\documentstyle[aps]{revtex}
\begin{document}
\def\la{\langle}
\def\ra{\rangle}
\def\om{\omega}
\def\o0{\omega_0}
\def\ov{\omega_V}
\def\os{\omega_s}
\def\O0{\Omega_0}
\def\Os{\Omega_s}
\def\Om{\Omega}
\def\do{\Delta\omega}
\def\vep{\varepsilon}
\newcommand{\beq}{\begin{equation}}
\newcommand{\eeq}{\end{equation}}
\newcommand{\beqa}{\begin{eqnarray}}
\newcommand{\eeqa}{\end{eqnarray}}
\newcommand{\w}{\omega}
\newcommand{\W}{\Omega}

\begin{title}
{\Large\bf Time dependence of evanescent 
quantum  waves.}
\end{title}
\author{\large J. G. Muga$^{1,2}$ and M. B\"uttiker$^1$}
\address{
${}^1$ D\'epartement de Physique Th\'eorique, Universit\'e de 
Gen\`eve, CH-1211, Gen\`eve 4, Switzerland\\ 
${}^2$ Departamento de Qu\'\i mica-F\'\i sica, 
Universidad del Pa\'\i s Vasco, Apdo. 644, Bilbao, Spain
}
\maketitle

\begin{abstract}
 The time dependence of 
quantum evanescent waves generated by a point source 
with an infinite or a limited frequency band is analyzed.
The evanescent wave is characterized by a forerunner (transient) 
related to the precise way the source is switched on. It is 
followed by an asymptotic, monochromatic wave 
which at long times reveals the oscillation 
frequency of the source. 
For a source with a sharp onset the forerunner 
is exponentially larger than the monochromatic solution
and a transition from the transient regime to the asymtotic regime 
occurs only at asymptotically large times. 
In this case, the traversal time for tunneling 
plays already a role only in the transient regime. 
To enhance the monochromatic solution compared to the 
forerunner we investigate (a)  frequency band limited sources
and (b) the short time Fourier analysis (the spectrogram)
corresponding to a detector which is frequency band limited. 
Neither of these two methods leads to a precise determination of the 
traversal time. However, if they are limited to determine the 
traversal time only with a precision of the traversal time itself 
both methods are successful: In this case the transient behavior 
of the evanescent waves is at a time of the order of the traversal 
time followed by a monochromatic wave which 
reveals the frequency of the source.

\end{abstract}

\section{Introduction}
In order to summarize essential aspects of the time dependence of
wave phenomena a number of
characteristic velocities or times have been defined. 
The {\em phase velocity}, $\om/k$, is the velocity of constant phase points 
in the stationary 
wave (assume $k>0$ for the time being)
\beq
e^{ikx-iwt}\,.
\eeq
The boundary conditions, the superposition principle and the
{\em dispersion relation}
$w=w(k)$ between the frequency $\omega$ and
the wavenumber $k$ determine the time evolution  
of the waves in a given medium.
When a group of waves is formed by superposition of stationary waves around a
particular $\omega$,
it propagates with the {\em group velocity} $d\om/dk$.
In {\em dispersive media} (where $w$ depends on $k$),
the group velocity can
be smaller (normal dispersion) or greater (anomalous dispersion)
than the phase velocity. It was soon understood that these velocities could
be both greater than $c$ for the propagation of light, 
and Sommerfeld and Brillouin \cite{Brill}, studying the fields that
result from an input step function modulated signal in a single Lorentz 
resonance medium, introduced other useful velocities, such as 
the velocity of the very first wavefront (equal to $c$),
or the {\em signal velocity}
for the propagation of the main front of the wave.  
Both the very first front and the signal velocity describe thus 
the causal response of the system and are therefore of particular interest. 
  
The above description is however problematic   
for {\em evanescent waves}, characterized by imaginary
wavenumbers instead of the real wavenumbers of propagating waves. 
Their time dependence has been investigated by
theoretitians and experimentalists in recent years because of its  
peculiar behaviour. A striking phenomenon   
is that, when crossing an evanescent region, certain initial wave
features (such as the peak of the incident 
amplitude) appear at the far side at anomalously large speeds, but 
clearly comparison of peaks of transmitted and incident wave packets
does not describe a causual process. 
Many publications and a recent workshop have been
devoted to discuss the implications \cite{workshop}.
         
The role played by the imaginary part of the 
group velocity $dw/dk$ and the possible definition of a signal velocity 
in the evanescent case
have been much discussed but not
yet completely clarified.
Assume that a source is placed at $x=0$ and emits with 
frequency $\o0$ from $t=0$ on. If $\o0$ is above the {\em cutoff frequency}
of the 
medium (the one that makes $k=0$)
a somewhat distorted but recognizable front propagates with
the velocity corresponding to $\o0$. Within the framework of the 
Schr\"odinger equation,
and using a set of dimensionless quantities where the cutoff
frequency is 1 (see the subsection below),
the dispersion relation takes the form \cite{qm}
\beq\label{disp}
\om=1+k^2\,, 
\eeq
and the signal propagation velocity for the main front is
equal to the group velocity,     
$v_p=(d\om/dk)_{\o0}=2(\o0-1)^{1/2}$. In other words, at some
distance $x$ form the source, the amplitude behaves, 
in first approximation, as
\beq
\psi(x,t)\approx e^{-i\o0 t}e^{+i k_0 x}\Theta(t-xv_p)\,
\eeq 
where $k_0=(\om_0-1)^{1/2}$ is the wavenumber related to  
$\o0$ by the dispersion relation, and $\Theta$ is the Heaviside (step) 
function.     
In the evanescent case, $\o0<1$, a preliminary analysis by Stevens 
\cite{Stevens}, following the contour deformation techniques 
used  by Brillouin and Sommerfeld
suggested that a main front,
moving now with velocity $v_m=2(1-\o0)^{1/2}={\rm Im}(d\om/dk)_{\o0}$,
and attenuated exponentially 
by $\exp(\kappa_0 x)$ (where $\kappa_0=(1-\o0)^{1/2}$),
could be also identified,  
\beq\label{front}
\psi(x,t)\approx e^{-i\o0 t}e^{-\kappa_0 x}\Theta(t-xv_m)\,.
\eeq 
The result seemed to be supported by a different approximate 
analysis of Moretti based on the exact solution \cite{MorPS},
and by the fact that    
the time of arrival of the evanescent front, 
$\tau=x/v_m$, had been found independently by B\"uttiker and Landauer
\cite{BuLa,Bu} as a characteristic {\em traversal time}
for tunnelling
using rather different criteria (semiclassical arguments, 
the rotation of the electron spin in a weak magnetic field,
and the transition from 
adiabatic to sudden regimes in an oscillating potential barrier).  

In the treatment of Stevens, as well as in the original work by 
Sommerfeld and Brillouin, the contour for the integral defining the 
field evolution was deformed along the
steepest descent path from the saddle point; and  
the main front (\ref{front}) was associated with  a residue due to
the crossing of a pole at $i\kappa_0$ by the
steepest descent path.  
But later, more accurate studies of the punctual source problem or other 
boundary conditions showed that 
the contribution from the saddle point 
(due to to frequency components above or at the frequency
cutoff created by the sharp onset of the source emission),
and possibly from other critical points (e.g. resonance poles
when a square barrier is located in front of the source \cite{BM})
were generally dominant at $\tau$, so that no sign of
the $\o0-$front (\ref{front}) can in fact be seen in the total wave
density at that time \cite{RMFP,RMA,TKF,APJ,BM}.
Similarly, corrections to the original work by Sommerfeld and 
Brillouin have been also worked out for electromagnetic pulse 
propagation \cite{OS}.  
In spite of these clarifying works, several important aspects have
remained obscure or not investigated, such as the  
actual time scale for the attainment 
of the stationary regime, the characterization of the
transients, and the role (if any) played by 
$\tau$ in the time dependence of the quantum wave.

Recently, one of the authors in collaboration with H. Thomas \cite{BT}, 
reconsidered the problem of Sommerfeld and Brillouin,
and provided a detailed discussion of the forerunners 
and the signal sent out by a source which has a sharp
onset in time. These authors also 
pointed out that the forerunner, generated by switching on the source, 
is associated with a time-dependent wide band spectrum 
whereas the signal, the $\o0-$front, 
carries the oscillation frequency of 
the source into the evanescent medium. The signal is called
a "monochromatic front". In contrast to previous work, which tried to 
find a front simply by analyzing the amplitude of the waves, these
authors emphasized the frequency content of the forerunner and 
the signal.
In the evanescent case, the amplitude of the monochromatic 
front is exponentially small compared to the forerunner, and in agreement 
with the works mentioned above, it cannot be detected using a simple 
criterion based on the magnitude of the wave.  
Two approaches were proposed to enhance the monochromatic fronts 
compared to the forerunners. First, the dominance of the forerunners
might arise due to the fact that high frequencies are transmitted 
in the propagating energy range. This can be avoided if the 
source is frequency limited such that all frequencies of the source 
are within the evanescent case. (Technically this means that a frequency
window is chosen to avoid the effect of the saddle point contribution.) 
A second option is not to limit the 
source but to frequency limit the detection. We can chose a detector 
that is tuned to the frequency of the source and that responds when 
the monochromatic front arrives.

The aim of this work is to characterize the time dependence 
of Schr\"odinger evanescent waves generated by a point source. 
We identify several wave features, in particular the arrival of 
the first main peak and the transition from a forerunner
dominated behavior to 
an asymptotic behavior dominated by the monochromatic front. We also 
investigate in some detail the proposals made in Ref.\cite{BT} to enhance 
the monochromatic front and consider both frequency limited sources and 
a frequency-time analysis of the wave at a fixed position. This leads 
to the investigation of the spectrogram of the wave generated by the 
source.

For a source with a sharp onset, 
we find that the traversal time $\tau$ plays a basic and  
unexpected role in the transient regime. 
For strongly attenuating conditions 
(in the WKB-limit) the traversal time governs the appearance 
of the first main peak of the forerunner. In contrast, 
the transition from the forerunner to an asymptotic regime 
which is dominated by the monochromatic signal of the source 
is given by an exponentially long time.
If the source is frequency band limited such that it switches on gradually
but still fast compared to the traversal time, the situation 
remains much the same as for the sharp source, except that 
now the transition fom the transient regime to the stationary 
regime occurs much faster, but still on an exponentially long time-scale. 
The situation changes if we permit the source to be switched on 
on a time scale comparable to or larger than the traversal time for
tunneling. Clearly, in this case a precise definition of the traversal 
time is not possible. But for such a source the transtion 
from the transient regime to the asymptotic regime 
is now determined by the traversal time. 
Much the same picture emerges if we limit instead of the 
source the detector. As long as the frequency window of the detector 
is made sharp enough to determine the traversal time with accuracy, 
the detector response is dominated by the uppermost frequencies. 
In contrast if the frequency window of the detector is made so narrow 
that the possible uncertainty in the determination of the traversal 
time is of the order of the traversal time itself, the detector 
sees a crossover from the transient regime to the monochromatic 
asymtotic regime at a time determined by the traversal time. 

Possibly, the fact that we can not determine the traversal time 
with an accuracy  better than the traversal time itself
tells us something fundamental about the tunneling time problem 
and is not a property of the two particular methods investigated
here. 

\subsection{Dimensionless quantities and notation}
The (dimensional) time dependent Schr\"odinger equation for a
particle of mass $m$ moving in a constant potential $V(X)=V$ 
is given by  
\beq
i\hbar\frac{\partial{\Psi}}{\partial T}=-\frac{\hbar^2}{2m}
\frac{\partial^2\Psi}{\partial X^2}+V\Psi
\eeq
The number of variables and parameters may be reduced by 
introducing dimensionless quantities for position, time and wave
amplitude, 
\beqa
x&=&\frac{X(2mV)^{1/2}}{\hbar}\\
t&=&\frac{TV}{\hbar}\\
\psi(x,t)&=&\frac{\hbar^{1/2}}{(2mV)^{1/4}}\Psi(X,T)\,.
\eeqa
This allows to write the corresponding dimensionless
Schr\"odinger equation,  
\beq
i\frac{\partial{\psi}}{\partial t}=
-\frac{\partial^2\psi}{\partial x^2}+\psi\,.
\eeq
Other useful dimensionless variables related to the
dimensional energy $E$ (or frequency 
$W=E/\hbar$), and wavenumber, $K=[2m(W-V/\hbar)/\hbar]^{1/2}$, 
are respectively  
\beqa
\omega&=&E/V\\
k&=&(\omega-1)^{1/2}=\frac{K\hbar}{(2mV)^{1/2}}\,.
\eeqa 
The reader may check that the dimensional dispersion relation 
\beq
W=\frac{V}{\hbar}+\frac{K^2\hbar}{2m}
\eeq
takes for dimensionless quantities the simple form 
given in (\ref{disp}). 
\section{Source with a sharp onset (Infinite frequency band)}
In this section we shall investigate the time dependent
wave function corresponding to the 
``boundary condition''
\beq\label{initt}
\psi(x=0,t)=e^{-i\o0 t}\Theta(t)\,,
\eeq
which may also be given by the corresponding Fourier transform 
\beq\label{init}
\widehat{\psi}(x=0,\om)=\frac{1}{(2\pi)^{1/2}}\frac{i}{\om-\o0+i0}\,.
\eeq
The superposition  
\beq
\psi(x,t)=\frac{i}{2\pi}\int_{-\infty}^{\infty}
d\om\, \frac{e^{ikx-i\om t}}{\om-\o0+i0}
\eeq
satisfies the Schr\"odinger equation as well as the 
boundary condition (\ref{initt}). 
Along the integration path, $k$ is positive for $\om>0$, and 
purely imaginary (with positive imaginary part) for $\om<0$. 
This corresponds to outgoing waves from the source. 
In Ref. \cite{BT} the integration was carried out in the complex 
frequency plane since this permitted a close comparison between 
the calculation for the Schr\"odinger equation and for relativistic
field equations. If only the Schr\"odinger equation is of interest 
the complex $k$-plane is most advantageous. In fact this permits 
to express the integral in terms of known functions. 
In the complex $k$-plane we have, 
\beqa
\psi(x,t)&=&-\frac{1}{2i \pi}\int_{\Gamma_+} dk\, 2k
\frac{e^{ikx-i(k^2+1)t}}{k^2+\kappa_0^2}
\\
\label{two}
&=&-\frac{e^{-it}}{2\pi i}\int_{\Gamma_+} dk
\left[\frac{1}{k+i\kappa_0}+\frac{1}{k-i\kappa_0}\right]
e^{ikx-ik^2t}\,,
\eeqa
where the contour $\Gamma_+$ goes from $-\infty$ to $\infty$
passing above the pole at $i\kappa_0$. 
The two terms in (\ref{two}) lead to integrals with the form discussed 
in the Appendix A. The contour can be deformed   
along the steepest descent path from the saddle at $k_s=x/2t$, 
the straight line
\beq
k_I=-k_R+x/2t\, ,
\eeq
($k_R$ and $k_I$ are the real and imaginary parts of $k$.) 
plus a small circle around 
the pole
at $i\kappa_0$ after it has been crossed by the steepest descent path, 
for fixed $x$, at the 
critical time  
\beq
\tau=\frac{x}{2\kappa_0}\,.
\eeq
This procedure allows to  
recognize two $w$-functions \cite{AS,FT} (see the Appendixes A and B),
one for each integral,   
\beq\label{exact}
\psi(x,t)=\frac{1}{2}e^{-it+ik_s^2 t}\left[
w(-u_0')+w(-u_0'')\right]\,.
\eeq
Here,  
\beqa\label{us}
u_0'&=&\frac{1+i}{2^{1/2}}t^{1/2}\kappa_0\left(-i-\frac{\tau}{t}\right)
\\
\nonumber
u_0''&=&\frac{1+i}{2^{1/2}}t^{1/2}\kappa_0\left(i-\frac{\tau}{t}\right)\,.
\eeqa
It is clear from the exact result (\ref{exact},\ref{us}),
that $\tau$ is an important parameter that appears naturally in
the $w$-function arguments, and determines with $\kappa_0$
the global properties of the solution. Its detailed role will be discussed 
in the following sections.   

Eq. (\ref{exact}) is in agreement with a previous expression by Moretti, 
derived using different contour deformations and notation. 
Our analysis of this exact result will be however
quite different and more detailed.  
\subsection{Approximations}    
Often an exact expression is not very informative by itself, and the  
approximations make its essential content manifest in 
certain limits.  
The simplest approximation for $\psi(x,t)$ for times before $\tau$
is to retain the
dominant contribution of the saddle
[by putting $k=k_s$ in the denominators of 
(\ref{two}) and integrating along the steepest descent path]. 
It is useful to write this in different ways,
\beqa
\psi_s(x,t)&=&\frac{e^{-it+ik_s^2t}}{2i\pi^{1/2}}\left(\frac{1}{u_0'}
+\frac{1}{u_0''}\right)
=\frac{e^{-it+ik_s^2t} \tau(2t/\pi)^{1/2}}{(i-1)\kappa_0(\tau^2+t^2)}
\nonumber\\
\label{psis}
&=&\frac{i}{2\pi}\sqrt{\frac{-4\pi i}{\Os t}}\frac{\Os}{\Os-\O0}
e^{-i(1-\Os )t}\,,
\eeqa
where
\beq
\Os\equiv k_s^2=\frac{x^2}{4t^2},\;\;\;\;\;\;\;\O0=-\kappa_0^2
\eeq
(These are particular values of the frequency variable $\Omega=w-1$,
defined with respect to the frequency level of the potential.)
The average local instantaneous frequency for this saddle 
contribution \cite {avw} is equal to the frequency of the saddle 
point,   
\beq\label{ws}
\os\equiv 1+\Os\,. 
\eeq
(Note the different sign of $\Os$
in the exponent of 
Eq. (\ref{psis}). $\Os$ depends on $t^{-2}$ and a time derivative has to be
taken to obtain (\ref{ws}), see \cite{avw}.)   
   
After the crossing of the pole $i\kappa_0$ by the steepest descent 
path at $t=\tau$ the residue 
\beq\label{resi}
\psi_0(x,t)=e^{-i\o0 t}e^{-\kappa_0 x}\Theta(t-\tau) .
\eeq 
has to be added to (\ref{psis}), 
\beq\label{apro}
\psi(x,t)\approx \psi_s(x,t)+\psi_0(x,t)\,.
\eeq
The solution given by Eq. (\ref{resi}) describes a monochromatic front 
which carries the signal into the evanescent medium. 
The conditions of validity of this  
approximation can be determined by examining the asymptotic
series of the $w(z)$ functions in (\ref{exact}) for large $|z|$, 
see the Appendix B. In fact (\ref{apro}) is obtained from the dominant 
terms of these expansions.   
The modulus of the two $w-$function arguments in (\ref{us})
is given by 
\beq
M(t)\equiv |-u_0'|=|-u_0''|=\frac{\kappa_0}{t^{1/2}}(t^2+\tau^2)^{1/2}\,.
\eeq
This quantity may be large in different circumstances.
It goes to $\infty$ when $\kappa_0$, $t$, or $x$ go to $\infty$, and also
when $t\to 0$. 
$M(t)$ is minimum at the crossing time $t=\tau$, where it takes the value
$M(\tau)=(x\kappa_0)^{1/2}$. For $x\kappa<1$ the pole lies within the range 
of the saddle Gaussian and cannot be treated separately.       

Assumming that $M$ is large the phase of $z$ determines the appropriate 
asymptotic expression of $w(z)$. For $Im(z)>0$, $w(z)\sim i/(\pi^{1/2}z)$,
whereas for $Im(z)<0$, $w(z)\sim i/(\pi^{1/2}z)+e^{-z^2}$,
see (\ref{wser}) and (\ref{zle}).       
${\rm arg}(-u_0')$ goes from $\pi/4$ to 
$3\pi/4$ when $t$ goes from $t=0$ to $\infty$, 
and ${\rm arg}(-u_0'')$ goes
from $\pi/4$, at $t=0$, clockwise to $-\pi/4$ when 
$t\to\infty$, and crosses the real axis at $t=\tau$. Thus 
the application of the previous asymptotic formulae lead exactly 
to Eq. (\ref{apro}). 
As discussed in the Appendix B one should not be 
mislead by the seeming front in (\ref{apro}). 
It is possible to obtain a smooth 
approximate expression around $t=\tau$ by adding a correction
to $\psi_s$ that takes into account the region near the pole \cite{BT}.  
This corrected expression however is asymptotically equivalent to (\ref{apro})
because at $\tau$, and 
within the conditions that make the saddle approximation valid, the
contribution of the pole is negligible. 
To see this more precisely 
let us examine the ratio between the 
modulus of the two contributions,  
\beq
R(t)\equiv\frac{|\psi_0|}{|\psi_s|}=
\frac{2\pi^{1/2}}{x} e^{-\kappa_0 x}t^{3/2}(x^2/4t^2+\kappa_0^2)\,.
\eeq
Its value at $\tau$ is an exponentially small quantity, 
\beq\label{Rt0}
R(\tau)=e^{-\kappa_0 x}(2\pi\kappa_0 x)^{1/2}\,.
\eeq
Note also that $R$ has a minimum at $\tau/3^{1/2}<\tau$ and
tends to $\infty$ as $t\to\infty$. As a function of $x$, the minimum 
of $R$ decreases up to $x\kappa_0=1/2$, and then grows again
monotonously.  

In summary, for the source 
with a sharp onset described here, the monochromatic front is not 
visible when the approximation 
(\ref{apro}) remains valid around $t=\tau$.  
However two very important observable features of the wave can be
extracted easily from (\ref{apro}). 
The first one is the arrival of the {\em transient front}, characterized 
by its maximum density at $t_f\equiv\tau/3^{1/2}$.
It is important to emphasize that this
time is of the order of $\tau$, but the wave front that arrives does not 
oscillate with the pole frequency $\o0$, but with the saddle point 
frequency $\os$. 
Figure \ref{f1} shows this transient front for three positions.
In this and similar figures the densities are exponentially amplified 
by $e^{2\kappa_0 x}$ to make possible the comparison among 
different values of $x$ with the same scale.
Thus, all amplified 
densities $A\equiv|\psi|^2 e^{2\kappa_0 x}$
tend to one in the asymptotic large $t$ regime,  and the
corresponding logarithm to zero.
Also shown is the approximation due to $\psi_s$, althout it is only 
distinguishable from the exact result for  $x=7$. Note that the amplified  
density $|\psi_s|^2 e^{2\kappa_0 x}$ is simply $R^{-1}$. 
In Figure \ref{f2} the instantaneous average frequency 
$\bar{w}$ is represented for the smallest $x$ value of Figure \ref{f1}, 
$x=7$.
At small $t$ and around the transient front,
$\bar{w}=\bar\simeq\om_s$.  
        
The second observable feature that we can extract from (\ref{apro})
is the time scale for the 
attainment of the stationary regime, or equivalently, the duration 
$t_{tr}$ of the transient regime dominated by the saddle before the 
pole dominates. $t_{tr}$ can be identified formally as the time where
the saddle and pole contributions
are equal, $R=1$. Because of 
(\ref{Rt0}) we shall assume  $\tau<<t_{tr}$ to obtain
the explicit result 
\beq\label{ttr}
t_{tr}\approx
\left(\frac{x e^{\kappa_0 x}}{2\kappa_0^2\pi^{1/2}}\right)^{2/3}\,.
\eeq
Using $\kappa_0 x>>1$,  the
velocity for the motion of the   
the space ``point'' where the transition from transient to 
stationary behaviour (or from saddle to pole) takes place is given by
\beq\label{vtr}
v_{tr}\approx\frac{3}{2\kappa_0 t}\,.
\eeq
Contrary to the motion of the front maximum, this point does not move with 
constant velocity. The transition may be observed in various ways.
In Fig. \ref{f1} the logarithms of the exact 
densities and of  their components are represented. The 
crossing point where $R=1$ may be observed for $x=7$ in the
change of behaviour of the total wave density.
The oscillation around $t_{tr}$ is due to the interference
between $\psi_0$ and $\psi_s$, and has
the characteristic period $T_0\equiv 2\pi/|\O0|$.
In Figure \ref{f2} the crossover
between the regime dominated by the saddle frequency and the one 
dominated by $\o0$ is also easily noticeable.

Finally, when $x\kappa_0$ is small ($\stackrel{<}{\sim}1$),
the saddle approximation  describes correctly 
the very short time initial growth, but fails around $\tau$  
because the pole is within the width of the Gaussian
centered at the saddle point. The pole cancels part of the 
Gaussian contribution so that the bump predicted by 
$\psi_s$ at $\tau/3^{1/2}$ is not seen in this regime. 
Fig. \ref{f3} shows the time dependence of the density 
for two small values of $x$. $\tau$ does not correspond to any 
sharply defined feature, but provides here a valid rough estimate of the 
attainment of the stationary regime.

\section{Frequency-band limited source}
H. Thomas and one of the authors \cite{BT} suggested recently to avoid 
the dominace of the 
saddle point solution over the 
monochromatic front by limiting the frequency band of the 
source. Specifically, it was proposed to cut-off  
the $\om$-amplitude (\ref{init}),  
\beq\label{cut}
\widehat{\psi}(w,x=0)=\frac{1}{(2\pi)^{1/2}}
\frac{i}{\om-\o0+i0}\left[\Theta(\om-(\o0-\do))
-\Theta(\om-(\om-\do))\right]\,.
\eeq
This implies that the emission of the source is not sharply defined 
as in (\ref{init}), see the Appendix C.  Instead, 
\beq
\psi(x=0,t)
=e^{-i\o0 t}\left[\Theta(t)-\frac{1}{\pi}{\rm Im} E_1(-i\do t)\right]\,,
\eeq 
where $E_1$ is the exponential integral defined in (\ref{exin}).  
The frequency band width $\do$ may be chosen so that
the onset of the source is fast with respect to $\tau$ 
and that all the frequencies in the source are in the evanescent 
region \cite{BT},
\beq\label{con}
\frac{2\pi}{\tau}\ll \do \ll |\O0|\,.
\eeq
Later we will see that in fact it is necessairy to give up the condition 
that the source switches on fast compared to the traversal time. 
Now ``negative times'' are also required to 
describe the signal growth.  
For an arbitrary $x$,  
\beq\label{fbl}
\psi(x,t)=\frac{i}{2\pi}\int_{\o0-\do}^{\o0+\do} d\om\,\frac{e^{-i(\om t-kx)}}
{\om-\o0+i0}\,.
\eeq
In this case the integration technique used
in the previous section does not provide an
analytical solution. We shall manipulate the integral 
to facilitate the exact numerical evaluation and the discussion of
approximations. 
Using $\Omega=\om-1$, Eq.(\ref{fbl}) becomes
\beq\label{37}
\psi(x,t)=\frac{ie^{-it}}{2\pi}
\int_{\O0-\do}^{\O0+\do} d\Om\,\frac{e^{-i(\Om t-kx)}}
{\Om-\O0+i0}\,.
\eeq
The integrand of Eq. (\ref{37}) has a saddle point 
at $\Os=(x/2t)^2$. 
But since $\Omega_0$ is now negative, a frequency interval 
chosen according to Eq. (\ref{con}), now excludes this saddle. 
It is convenient to
introduce the variable $y=\Om/\Os$ so that the valley-hill structure 
of the exponent remains constant, 
\beq\label{inty}
\psi(x,t)=\frac{ie^{-it}}{2\pi}\int_{y_-}^{y_+} dy\,
\frac{e^{i\lambda {\rm sign}(t)(2y^{1/2}-y)}}{y-y_0+i0}\,.
\eeq
Here, 
\beq
y_0=\O0/\Os\,,\;\;\;\;y_{\pm}=y_0\pm \do/\Os\,,
\;\;\;\;\lambda=|\Os t|\,.
\eeq
In order to perform contour deformations in the complex $y$ plane
it is also 
necessary to specify that the branch cut for $y^{1/2}$ is set along 
the positive real axis, and that  
the saddle is at $y=1$ for $t>0$ (just above the cut) but at 
$y=e^{2i\pi}$ for $t<0$ (just below). The steepest descent and ascent paths 
from it are given, in the first Rieman sheet of $y$,
by the sections of the parabolas 
\beqa
y_I&=&{\rm sign}(t) \frac{1}{2}(1-y_R)^2\;\;\;\;\;\; ({\rm descent})
\\
y_I&=&-{\rm sign}(t) \frac{1}{2}(1-y_R)^2\;\;\;\;\;\; ({\rm ascent})
\eeqa
starting from the saddle point. 
Irrespective of the sign of $t$ the original contour 
in (\ref{inty}) is entirely within one of the valleys of the saddle, 
so that there is no need to take this critical point into account in the
contour deformation.  The most efficient contour deformation
consists on following the 
steepest descent path from the lower integration extreme $y_-$
downwards to infinity and coming up to the upper integration extreme  
$y_+$ following 
its steepest descent path. If $t>0$ this contour encloses the pole 
at $\o0$ and the corresponding residue has to be included. The extreme 
points become the critical points of the integral, apart from the pole 
at $\o0$ when $t>0$,  
\beq
\psi(x,t)=D_- -D_+ +\psi_0'\,,
\eeq
where $D_\pm$ (generically $D_z$) are the integrals from $y_\pm$
to $\infty$
along the corresponding steepest descent paths, $SDP(z)$,
\beq\label{Dz}
D_z=\frac{ie^{-it}}{2\pi}\int_{SDP(z)} dy\,
\frac{e^{i\lambda{\rm sign}(t)(2y^{1/2}-y)}}{y-y_0}\,,
\eeq
and 
\beq
\psi_0'(x,t)=e^{-i\o0 t}e^{-\kappa_0 x}\Theta(t)\,.
\eeq
Unlike (\ref{resi}) the residue is now present at all positive times.  
Using the generic notation $z=y_\pm$ for any of the two extreme points 
the steepest descent paths from them and their slopes are given by  
\beqa
y_I&=&\frac{1}{2}{\rm{sign}}(t) [y_R^4-4y_R^3(1+z)+2y_R^2 z (4+3z)
-4z^2y_R(1+z)+z^4]^{1/2}
\\
\frac{\partial y_I}{\partial y_R}&=&
\frac{y_R^3-3y_R^2(1+z)+y_R z(4+3z)-z^2(1+z)}{2y_I}
\\
\frac{\partial y_I}{\partial y_R}\Big|_z&=&{\rm sign}(t)(-z)^{1/2}\, .
\eeqa
In the numerical evaluations
the value of $y$ along the path may 
be obtained by solving for each step  
\beq
dy=dy_I(dy_R/dy_I+i)\,,
\eeq
with the initial condition $y=z$.
\subsection{Approximations}
The leading term in $\lambda^{-1}$ of (\ref{Dz}) is obtained by
integrating from $z$ along the ray with the direction of the 
steepest descent path at $z$ and taking all functions
in the integrand, except the exponential,
out of the integral, with their values at $z$ \cite{BH},   
\beq\label{Dza}
D_z\sim D_z^0\equiv \frac{i}{2\pi \lambda (z-y_0)|z^{-1/2}-1|}
e^{-it\omega_z-\kappa_z x}
e^{i\theta_z}\,,
\eeq
where
$\omega_z=\omega_\pm=\o0\pm\do$,
$\kappa_z=\kappa_\pm=(1-\omega_\pm)^{1/2}$,
and $\theta_z$ is the angle of
the steepest descent path at $z$,
\beq
\theta_z=\pi+\arctan[{\rm sign}(t)(-z)^{1/2}]  
\eeq
(the branch of the $\arctan$ function between
$-\pi/2$ and $\pi/2$ is taken).
The modulus of these contributions is given by
\beq
|D_\pm|=\frac{e^{-x\kappa_\pm}}{2\pi\do(t^2+t_\pm^2)^{1/2}}\,,
\eeq
where
\beq
t_\pm=\frac{x}{2\kappa_\pm}
\eeq
are the critical times when the steepest descent path from the 
saddle crosses the extreme points of integration $y_\pm$.
Regarding the possibility of making the monochromatic front visible, 
the frequency band limited source elliminates the saddle 
dominated transient effects, but they are substituted by transients 
associated with a new critical point:  the upper extreme of integration. 
The actual time for transition to the stationary regime is still 
larger than $\tau$ under semiclassical conditions, but much smaller than 
for the sharp onset case. 
A difference with the infinite frequency band case is that     
the maximum of $|D_\pm|$ is at $t=0$.
This is quite remarkable. A particular 
feature of the wave, its maximum, appears instantly with zero delay 
at arbitrarily large distances. One should keep in mind though that 
the build up of the initial signal has required previously an
infinite time, from $t=-\infty$ to $t=0$.    

The transient regime previous to the dominance of the pole must be also 
different from the infinite band case since now there is no
saddle point contribution. The transient 
is here dominated by the integral $D_+$ because of the exponential 
dependencies and 
by the frequency of the upper integration limit $y_+$.
An analysis similar to the one performed in the infinite frequency 
band case can be now performed. Let us compare     
the contributions from the pole and $D_+$ by dividing their 
moduli, 
\beq\label{rprim}
R'=\frac{|\psi_0'|}{|D_+|}=
e^{(-\kappa_0+\kappa_+)x} [2\pi\do (t^2+t_+^2)^{1/2}]\,.
\eeq
From the condition $R'=1$ we obtain the new critical time
\beq\label{ttrp}
t'_{tr}=\left(\frac{e^{2x(\kappa_0-\kappa_+)}}
{4\pi^2\do^2}-t_+^2\right)^{1/2}
\approx \frac{e^{x(\kappa_0-\kappa_+)}}{2\pi\do}\,,
\eeq
and the critical velocity
\beq\label{vtr'}
v'_{tr}\approx \frac{1}{t(\kappa_0-\kappa_+)}\,.
\eeq
They have to be compared with the corresponding quantities 
for the sharp onset case, (\ref{ttr}) and (\ref{vtr}) respectively. 
The critical time for the transition depends exponentially on
$x$ in both cases, but it arrives exponentially
earlier for the frequency band limited case. 
In Figure \ref{f4} the logarithm
of the amplified density and of the approximation 
provided by $|D_+^0|^2$ is represented 
for two values of $x$. For $x=50$ the transition time $t_{tr}$ is relatively 
small and can be seen in the time interval shown (The oscillation period
of the interferences around
$t_{tr}$ is now $2\pi/\do$). For $x=135$,  $t_{tr}$ is much larger and 
cannot be seen in the scale chosen; note the well defined peak at 
$t=0$. The slight disagreement between 
the exact result and the approximation provided by $D_+^0$ is because, 
for very small times, the slope of the steepest descent path
from $y_+$ is very small and the path passes
close to the pole. The pole perturbs the 
integral in this manner for times such that
$\lambda^{-1}<\do/\Os$, i.e., $t<2\pi/\do$.

From Eq. (\ref{rprim}) we see that the dominance of the 
monochromatic front at $\tau$ and fixed $x$ requires that the 
magnitude of $(-\kappa_0+\kappa_+)x \approx 1$ or smaller.
This can be achieved in two ways: (a) We can consider a frequency 
$\omega_0$, and a frequency 
interval such that  $-\kappa_0$ and $\kappa_+$
become small. This means that both $1-\o0$ and $\do$ should be close to zero, 
but to avoid propagating waves $\do<(1-\o0)$;
(b) Alternatively, we can allow $\o0$ to take any other value below 1, 
but making the
the frequency interval $\Delta \omega$ small.
In both cases a small frequency window is needed and we can
expand with respect to $\do$ so that   
$(-\kappa_0+\kappa_+)x = - \Delta \omega \tau$.
Thus we see that the attempt 
to measure the traversal time accurately has to be abandoned.
The lower limit of Eq. (\ref{con}) is now violated. 
Figure \ref{f5} shows an example where the first inequality in (\ref{con}) is
not obeyed. The traversal time (rather than $t_{tr}$, which in this 
case is an imaginary number) now marks the transition to the
asymptotic region even though the build up time is so large that 
the arrival of the monochromatic front cannot be defined with a precission 
better than $\tau$ itself. In this respect, 
Figure \ref{f6} is very illustrative. It shows how the transition of the
average frequency characterizing the 
transient to the frequency of the monochromatic solution occurs  
around the traversal time $\tau$, but the time interval required for the 
transition is clearly larger than $\tau$.  
\section{Time-frequency analysis of the wave function}  
Instead of limiting the source in frequency we can limit the 
detection of the field to a range of frequencies of interest [14]. 
The wave amplitude at fixed position becomes a function of time,
$\psi(t)$,   
that can be Fourier analyzed to provide its frequency representation, 
$\widehat{\psi}(\om)$. 
However, neither of the corresponding densities, $|\psi(t)|^2$
or $|\widehat{\psi}(\om)|^2$, tells when a particular 
frequency arrives or decays, nor what is the relative importance, at a
given time, of different frequency components. This type of information is  
provided by joint time-frequency representations \cite{Cohenb}.
There are many possible ways to carry out a time-frequency
analysis. A simple one, and surely the most common, is the ``spectrogram''
based on the short time Fourier transform (stFt).
We are interested in the 
Fourier spectrum that would be measured at the observation 
point $x$ if the wave is observed during a time 
interval of duration $T$. 
If a precise determinantion of the traversal time is attempted, 
the extend of the time interval $T$ is chosen to be short compared 
to the traversal time $\tau$. On the other hand,
if $T$ is too short then the frequency resolution will be poor. 
A compromise to obtain a good resolution in time and frequency is for  
$T$ to be bound above and below [14],    
\begin{equation}
\frac{2\kappa_0}{x}=
\frac{1}{\tau} \ll \frac{1}{T} \ll 
\frac{|\Om_0 |}{2\pi}=\frac{\kappa_0^2}{2\pi}.
\label{Tcond}
\end{equation}
This is equivalently expressed in terms of inequalities for time scales 
\beq
\tau >> T >> T_0 , 
\eeq
where $T_0\equiv 2\pi/|\O0|$ is the oscillation
period corresponding to the frequency 
$|\O0|$.
Combining the inequalities one finds $\kappa_0 x>>1$, namely when these 
conditions are satisfied the saddle contribution to the wave function
will be a very good approximation to the total wave up to  
an ``exponentially long'' time $t_{tr}>>\tau$ (see section 2).     

The short time Fourier transform (stFt) of the field $\psi (x,t)$
is given by
\begin{equation}
F(\omega;x,t) =\frac{1}{(2\pi )^{1/2}} \int^{t+T/2}_{t-T/2} dt^{\prime}
\exp(i\omega t^{\prime}) \psi (x,t^{\prime}) .
\label{sF}
\end{equation}
Note that the short time Fourier spectrum depends, in addition 
to the frequency,  parametrically on the observation point $x$ and 
the time $t$.

Consider now first the contribution of the monochromatic front 
to the stFt. We denote this stFt by $F_{p}(\omega;x,t)$. 
For $t < \tau-T/2$ we have  $F_{p}(\omega;x,t) = 0$. For $t > \tau+T/2$  
we find 
\begin{equation}
F_{p}(\omega;x,t) =
\frac{2}{(2\pi )^{1/2}} e^{-\kappa_0 x} e^{i(\omega - \omega_{0})t}
\frac{\sin[(\omega - \omega_{0})T/2]}{(\omega - \omega_{0})}\,.
\label{sFp}
\end{equation}
This amplitude peaks at the frequency of the source $\omega_{0}$
and, for $t>\tau+T/2$, it is time-independent at this frequency.
Away from this frequency
the stFt of the monochromatic wave decays algebraically and 
oscillates sinusoidally. In the time interval $\tau -T/2 < t < \tau +T/2$
the stFt tracks the arrival of
the monochromatic front, 
\begin{equation}
F_{p}(\omega;x,t) =
\frac{ie^{-\kappa_0 x}}{(2\pi )^{1/2}}
\frac{1-e^{i(t+T/2)(\omega-\omega_0)}}{\om-\om_0}\,.
\end{equation} 
Consider next the saddle point contribution. Its frequency 
in a short time interval is determined by the expansion 
of $\Omega_s(t^{\prime}) t^{\prime}$ away from $t$. 
This expansion gives 
$\Omega_s(t^{\prime}) t^{\prime} = 
2\Omega_s (t) t - \Omega_s (t) t^{\prime} + O((t^{\prime}-t)^{2}).$ 
Taking into account that the prefactors 
are slowly varying over the time interval of interest
here (this is assured by the first inequality in (\ref{Tcond})),
we obtain, 
\begin{equation}\label{sFs}
F_{s}(\omega;x,t)  =  \frac{i}{\pi} 
\sqrt{\frac{-2 i}{\Om_s t }}\,
\frac{\Om_s}{\Om_s - \Om_0}
e^{i(\Om_{s} - 1 + \omega) t}
\frac{\sin[(\Os+1-\omega)T/2]}
{(\Os +1 -\omega )}\,.
\end{equation}
The stFt of the saddle point solution peaks at the frequency 
$\omega_s\equiv \Om_{s} + 1$.
This frequency is very large at 
short times (``kinetic regime''), where
$\om_s\sim \Om_s=x^2/(4t^2)$ 
is dominated by frequencies above the potential.  
At long
times (``potential regime'') $\om_s$ tends to  the potential frequency $1$. 
The transition time between these two regimes of different $\omega_s$
behaviour may be
estimated, by solving
$\Omega_s(t)=1$, as $t=x/2$ (it can be larger or smaller than 
$\tau$.)

We see that the two Fourier transforms peak at well separated
frequencies.
Still the stFt of the pole has an exponentially small amplitude
compared to that of the forerunner. Therefore, it is possible 
that the stFt of the saddle is still large at the frequency $\omega_0$
where the stFt of the monochromatic front peaks. Thus we have to 
investigate the stFt of the saddle at the frequency $\omega_0$.
From Eq. (\ref{sFs}) we obtain,
\begin{equation}\label{sFs0}
F_{s}(\omega_0;x,t)  =  \frac{i}{\pi} 
\sqrt{\frac{-2i}{\Om_s t }}\,
\frac{\Om_s}{\Om_s - \Om_0} 
e^{i(\Om_{s} + \Om_{0})t}
\frac{\sin[(\Om_{s} -\Om_{0})T/2]}
{(\Om_{s} -\Om_{0})} .
\end{equation}
At a time $t=\tau$ the amplitude of the saddle  is still
of order $1$. Thus  
even at the peak frequency of the front its contribution 
to the spectrum is of order $1$ and much larger than the exponentially
small peak of the monochromatic front. 

Let us now investigate the properties of the transient 
in the frequency-time domain, and determine the role 
played by $\tau$, in particular at the frequency 
of the signal $\om_0$.
The spectrogram is defined as the square modulus of the stFt,
$S(t,\omega; x) = N |F(\omega;x,t)|^{2}$, where $N$ is a normalization
constant. (The notation of the argument 
of $S$ is appropriate for a time-frequency analysis at fixed $x$.)
For ``normalizable'' cases $N$ is chosen so that $\int\int dt d\om S=1$.
In our case, this is not possible because the asymptotic  
stationary regime does not decay, but $S$ provides anyway information on  
the relative importance of two time-frequency points, so $N$
is chosen to be some convenient value, $N=\pi^2/2$.   
      
For analyzing the transient we may neglect the monochromatic
front and concentrate on the saddle point term. 
The spectrogram of the saddle point solution is denoted by 
$S_{s}(t,\omega;x)$ and is given by 
\begin{equation}
S_{s}(t,\omega;x) =  \alpha(t)
\frac{\sin^{2}[(\Om_{s} + 1 -\omega)T/2]}
{(\Om_{s} + 1 -\omega)^{2}} 
\label{Ss}
\end{equation}
with an amplitude
\begin{equation}
\alpha(t) = {\frac{1}{t}}\,
\frac{\Om_s}{(\Om_s - \Om_0)^{2}} .
\label{pSs}
\end{equation}
Note that $\alpha(t)$ is also proportional to the absolute square 
of the saddle point solution $|\psi_s(x,t)|^{2} = (4/\pi)
\alpha(t)$. The amplitude $\alpha$ has a maximum as a function
of time at the transient front peak, $t_f=3^{-{1/2}}\tau$,
that moves with a speed
\begin{equation}
v_{f} = \sqrt{3} v_m. 
\label{vmax}
\end{equation}
Thus $v_{f}$ determines both the speed of the peak value 
of the saddle point solution $|\psi_s(x,t)|^{2}$ and the speed of the 
absolute maximum of the spectrogram. The peak value of the spectrogram 
for a given time is at the frequency $\omega_s= \Om_{s} + 1$, 
see an example in Figure \ref{f7}. 
Consider next the spectrogram at the frequency of the source,
$\omega = \omega_0$. For fixed $x$ it is bounded by the envelop function 
\begin{equation}
\beta(t) = \frac{\Om_s}{t(\Om_s - \Om_0)^{4}} 
\label{aSs}
\end{equation}
and  has local maxima close to the times at which 
the $\sin$ function is $\pm 1$.
$\beta(t)$ grows as $t^5$ for short times and decays as $t^{-3}$ for large 
times. Its maximum arrives, for fixed $x$, at $\sqrt{5/3}\tau$. 
Equivalently, this maximum  moves 
with a speed
\begin{equation}
v_{env} = \sqrt{3/5}\, v_m . 
\label{vmin}
\end{equation}
This reveals that, in $S(t,\o0;x)$, 
$\tau$ 
is not the time where  
the stationary regime begins (it is necessary to 
wait an exponentially long time to attain that regime), 
but the basic time scale 
for the arrival of the main part of the transient. 
The detailed oscillatory pattern of this main part will depend on the 
value of $T/T_0$, see the Figures \ref{f8} and \ref{f9}.  
For $\om=\om_0$ the $\sin$ function vanishes at times
\beq\label{tn}
t_n=\frac{\tau}{(n T_0/T-1)} 
\eeq
(In the $\omega - t$ plane 
the spectrogram has a maximum which follows the line 
$\omega = \Om_{s} + 1$ and vanishes along 
the lines $\omega = \Om_{s} + 1+2\pi n/T$. In between 
these zero-lines the spectrogram exhibits local maxima.)  
The $n$ for the closest zero to $\tau$ is $n\approx 2T/T_0$ 
whereas the largest $t_n$ (associated with the
minimum $n$ so that (\ref{tn}) has real 
solution) corresponds to $n\approx T/T_0$. 
Hence there are approximately  $T/T_0$ zeros between 
$\tau$ and the last oscillation.       
For $T/T_0<1$ there is a single major bump close to $\tau$, and  
for $T/T_0>>1$ there are many local maxima that ``sample'' 
the form of the envelop function (\ref{aSs}).

\subsection{Other time-frequency distributions}
The time-frequency characterization of the wave function at fixed position 
depends on the quasi-distribution chosen. There is nothing wrong with 
this non-uniqueness as long as the quasi-distribution chosen is specified.
In choosing the distribution one may consider several factors.   
Ideally the distribution should be easy to calculate, and
should not be too noisy 
(e.g. with wild oscillations). It may also occur that 
one of the distributions is naturally adapted to the 
way the detection experiment is performed. 

Let us first discuss a different type of spectrogram. 
The roles of time and frequency can be inverted so that    
instead of limiting the Fourier transform of $\psi(t)$
with a ``square window''
\beq
h(t)=\Theta(t-T/2)\Theta(-t-T/2)\,, 
\eeq
a ``short frequency Fourier transform'' (sfFt) may be similarly defined
for $\widehat{\psi}(\om)$   
by limiting the frequency integral with a window,
\beq\label{gw}
g(w)=\Theta(\om-\Delta \om)\Theta(-w-\Delta \om) . 
\eeq
The corresponding spectrogram, $S'$, is obtained as the square modulus of 
the sfFt.  
The two spectrograms, $S$ and $S'$, are however not equal because
the window functions 
$h$ and $g$ are not Fourier transforms of each other. In particular, note
that 
the spectrogram $S'$ may be equivalently obtained from a short frequency  
Fourier transform analysis for an initially sharp onset signal,
or as the density that results from a source
with a smooth onset (The density of $\psi(x,t)$ in Eq. (\ref{fbl})
is proportional, up
to a trivial normalization constant, to $S'(t,\om_0;x)$.)
As a consequence, $S'(t,\om_0)$ peaks at $t=0$
after an infinite time growth whereas $S(t,\om_0)$ is strictly zero  
at $t<-T/2$ and its main transient part peaks around $\tau$.
The qualitative behaviour is different because of the different window
functions, even though the inequalities  used 
here or in section III, Eqs. (\ref{Tcond}) and (\ref{con}),
are in fact identical if one identifies 
$T=2\pi/\do$. (Compare Figure \ref{f8}
and the upper curve of Figure \ref{f4} where 
the values of $\tau$, $T$ and $T_0$ are equal.)                     
These inequalities imply in both cases that  
the pole contribution can be entirely neglected at or
around $\tau$, and that it will only be of importance at a
a time that depends exponentially on $\kappa_0 x$. 

We have also considered the time-frequency Wigner function.
This representation separates much more clearly the
saddle and pole contributions (it also associates to the later 
an initial time $t=0$ rather than $t=\tau$),
but presents the inconvenience of a very 
rapidly oscillating pattern and strong interference terms.

\section{Summary and discussion}
The time dependence of evanescent Schr\"odinger waves 
created by a point source has been investigated, combining
analytical or numerical exact results and approximate expressions. 
The background of the former allow to test and contrast the 
validity of the simplified description of the later.
We have also performed a time-frequency analysis of the 
transients which preceed the stationary regime by means of  
spectrograms.     

An important aspect of the work is the elucidation of the role played 
by the different parameters, in particular by the
``traversal'' time $\tau$.
It is a basic parameter that determines the global shape of the 
wave but at this time, in semiclassical conditions $\kappa_0 x>1$,
the monochromatic front 
(with the main frequency of the source) is not observed in the total wave 
because of the dominance of the saddle point contribution.  
However, for the source with a sharp onset we have identified a
transient front 
whose maximum peak arises at a time $\tau/3^{1/2}$. The dominant frequency 
of this front does not correspond to the main frequency of the
source, $\o0$,  but to 
the mean saddle point frequency (above or
at the cutoff frequency).
The time $t_{tr}>\tau$ that determines the duration of the transient, 
or the passage to the asymptotic regime, 
has also been identified.
Frequency band limited sources accelerate the crossover to the stationary 
regime but there is still a transient dominated by the upper frequency
contained in the band.
The spectrogram $S(t,\om;x)$ has been also investigated to show
the variation with time of the
frequency components.    
In this representation the main
contribution at the
frequency $\o0$ arrives at $x$ around the traversal time $\tau$. This 
main contribution is a transient due to the saddle point.
At an exponentially long time  
the pole term will eventually take over and remain as the dominant
contribution in the stationary regime.      

In fact it is possible to see the mochromatic wave front arriving around
the traversal time
$\tau$, but only when the semiclassical conditions are abandoned, 
and with an accuracy which is never better than the traversal time itself. 
This occurs if the point source starts the emission abruptly, but
also when it is switched on gradually (limiting the frequency band), 
or for a limited frequency band detector, whose response is modelled here
with a spectrogram. The coincidence of all these cases strongly suggests 
that the limitation of the accuracy to measure $\tau$ is in fact a general 
property.

\acknowledgments{We are indebted to Harry Thomas who through discussions during the work of Ref.\cite{BT} has also strongly influenced the direction of this
work. 
J. G. M. was supported by 
MEC (PB97-1492), and in 
Geneva, where most of this work was carried out, by 
the Swiss National Science Foundation. M.B. is 
supported by the Swiss National Science Foundation.}  

\appendix
\section{Integrals}
In this appendix we shall solve the integrals of the form  
\beq
{\cal I}=\int_{\Gamma_+}dk\, \frac{e^{-i(ak^2+kb)}}{k-k_0}\;\;\;\; a>0\,,
\eeq
where ${\Gamma_+}$ goes from $-\infty$ to $\infty$ passing {\it above} the
pole at $k=k_0$. The saddle point of the exponent is at $k=-b/2a$ and the 
steepest descent path is the straight line $k_I=-(k_R+b/2a)$. The original 
contour is at the border between hill and valley and can be deformed
into this path, taking into account the residue of the pole when 
${\rm Im}(k_0)<-[{\rm Re}(k_0)+b/2a]$.  By completing the square 
and introducing the new variable
\beq
u=u(k)=\frac{1+i}{2^{1/2}}a^{1/2}(k+b/2a)\,,
\eeq
the integral takes the form
\beq
{\cal I}=e^{i\frac{b^2}{4a}}\left\{\int_{-\infty}^{\infty}
\frac{e^{-u^2}}{u-u_0} du - 2i\pi e^{-u_0^2}\Theta[\rm{Im}(u_0)]\right\}\,
\eeq
where $u_0=u(k=k_0)$.
Note that the $u-$variable has its origin at the saddle point and its real
axis corresponds to the steepest descent line.  
Using (\ref{wint}) and (\ref{A3}) it can be finally written as 
\beq
{\cal I}=-i\pi e^{i\frac{b^2}{4a}} w(-u_0)\,.
\eeq
\section{Properties of $\lowercase{w(z)}$}
The $w-$function \cite{AS,FT} is an entire function defined in terms of
the complementary error function as
\beq
w(z)=e^{-z^2}{\rm erfc}(-iz)\,.
\eeq
$w(z)$ is frequently recognized by its integral expression  
\beq\label{wint}
w(z)=\frac{1}{i\pi}\int_{\Gamma_-}\frac{e^{-u^2}}{u-z}du
\eeq
where $\Gamma_-$ goes from $-\infty$ to $\infty$ passing below 
the pole at $z$. For ${\rm Im} z>0$ this corresponds to an integral
along the 
real axis. For ${\rm Im} z<0$ the contribution of the residue has to be
added, and for ${\rm Im} z=0$ the integral becomes the principal part 
contribution along the real axis plus half the residue. 
From (\ref{wint}) two important properties are deduced, 
\beqa\label{A3}
w(-z)&=&2e^{-z^2}-w(z)\\
\noalign{\hbox{\rm and}}
w(z^*)&=&[w(-z)]^*.
\eeqa
To obtain an asymptotic series 
as $z\to\infty$ for ${\rm Im} z>0$ one may expand  
$(u-z)^{-1}$ around the origin (the radius
of convergence is the distance
from the origin to the pole, $|z|$) and integrate term by term. This
provides   
\begin{equation}\label{wser}
w(z)\sim \frac{i}{\sqrt{\pi}\, z}\left[1+\sum_{m=1}^\infty
\frac{1\cdot 3\cdot ...\cdot (2m-1)}{\left(2z^2\right)^m}\right]
\,\;
{\rm Im} z>0
\end{equation}
which is a uniform expansion in the sector ${\rm Im} z>0$.
For the sector ${\rm Im} z<0$ (\ref{A3}) gives
\beq\label{zle}
w(z)\sim \frac{i}{\sqrt{\pi}\, z}\left[1+\sum_{m=1}^\infty
\frac{1\cdot 3\cdot ...\cdot (2m-1)}{\left(2z^2\right)^m}\right]
+2 e^{-z^2}  
\,,\;
{\rm Im} z<0\,.
\eeq
If $z$ is in one of the bisectors then $-z^2$ is purely imaginary 
and the exponential becomes dominant.
But right at the crossing of the real axis, ${\rm{Im}} z=0$,
the exponential term is of 
order $o (z^{-n})$, (all $n$), so that (\ref{wser}) and (\ref{zle}) are
asymptotically equivalent as $|z|\to\infty$. One may insist however
in removing the discontinuity of the series expansion (as a function of
$z$) around the 
real axis. This can be done by adding a correction to the saddle 
contribution that takes into account the region close to the pole
in the line integral, see also the discussion in Ref. \cite{BT}. 
Note that in (\ref{wser}) only the region around the origin
contributes. Consider now also the region around the pole, 
\beq
\int_{z_R-\Delta}^{z_R+\Delta} \frac{e^{-u^2}}{u-z}du
\approx e^{-z^2}\int_{z_R-\Delta}^{z_R+\Delta} \frac{du}{u-z}=
e^{-z^2}\ln\left(\frac{z_R+\Delta-z}{z_R-\Delta-z}\right)\,. 
\eeq
For large $|z|$ and in the proximity of the real axis this may be 
approximated by $e^{-z^2}i\pi{\rm{sign}}(z)$ which exactly 
cancels the discontinuity between (\ref{wser}) and (\ref{zle}).

\section{Time dependence of frequency band limited source}
The time dependence of the frequency band limited
amplitude at $x=0$ is obtained from (\ref{cut}) by Fourier transform
\beq\label{C1}
\psi(x=0,t)=\frac{1}{2\pi}
\int_{\o0-\do}^{\o0+\do} d\om\,\frac{i e^{-i\om t}}{\om-\o0+i0}
=\frac{1}{2}e^{-i\o0 t}+\frac{1}{2\pi} {\cal P}\int_{\o0-\do}^{\o0+\do}   
\frac{i e^{-i\om t}}{\om-\o0}\,. 
\eeq
Using the frequency $\om'=(\om-\o0)t$,
\beq
{\cal P} \int_{\o0-\do}^{\o0+\do} d\om\,   
\frac{e^{-i\om t}}{\om-\o0}=e^{-i\o0 t}{\cal P}
\int_{-t\do}^{t\do} d\om'\,\frac{e^{-i\om'}}{\om'}\,,
\eeq
and the principal part integral can be expressed by contour deformation 
in terms of combinations of exponential integrals \cite{JCP},
\beq\label{C3}
{\cal P}\int_{-b}^a dY\,\frac{e^{-iY}}{Y}=E_1(-ib)-E_1(ia)-i\pi\,,\;\;\;\;
a,b>0\,,
\eeq
where
\beq\label{exin}
E_1(z)=\int_z^\infty dY\,\frac{e^{-Y}}{Y},\;\;\;\;\;\;\;|{\rm arg} z|<\pi
\eeq
(the contour does not cross the negative real axis).  
Finally, using (\ref{C3}), (\ref{C1}) can be written, for all $t$, 
\beq
\psi(x=0,t)
=e^{-i\o0 t}\left\{\Theta(t)-\frac{1}{\pi}{\rm Im}[E_1(-i\do t)]\right\}.
\eeq
\section{Relativistic case}
Defining the dimensionless parameter $c$ by combining the dimensional
velocity of light $C$, the mass and the potential constant,  
as 
\beq
c=C(2m/V)^{1/2} ,
\eeq
the dispersion relation for a Klein-Gordon wave equation 
takes the form
\beq
\Om^2=(\om-1)^2=(c^2/2)^2+c^2k^2 .
\eeq
For evanescent conditions $k$ is purely imaginary, $k=i\kappa$.
In particular 
for $\om=\o0$, or $(\Om=\O0)$, $\kappa_0$ cannot take arbitrarily
large values,
$0<\kappa_0<c/2$, contrast this with the non-relativistic case where 
there is no upper limit.      
    
For a source with a sharp onset the wave function is \cite{BT}
\beq
\psi(x,t)=\frac{i}{2\pi}e^{-it}\int_{-\infty}^{\infty}
\frac{1}{\Om-\O0+i0}e^{-i(\Om t-kx)}\,d\Om ,
\eeq
which is zero for $t<x/c$.
There are now two saddle points at 
\beqa
\pm\Os&=&\pm \frac{c^2 t}{2\theta}
\\
\theta&\equiv& (t^2-x^2/c^2)^{1/2}
\eeqa
and two branch cuts.  
The integration contour may be deformed along the two steepest descent
paths.
The pole is crossed at \cite{BT} 
\beq
\tau=\frac{x}{2\kappa_0} .
\eeq
Note the lower bound $x/c<\tau$ when $\Om\to 0$ (or $\kappa_0\to c/2$).  
Consequently the traversal time is strictly limited by the velocity 
of light \cite{BT}. 
  
We shall only discuss the contribution from the saddle at  
$+\Os$ corresponding to the excitation of particles \cite{BT} , 
\beq\label{sr}
\psi_s^+(x,t)=\frac{ix}{c^2t}\left(\frac{-i}{\pi \theta}\right)^{1/2}
\frac{\Os}{\Os+\O0} e^{-i[t+\frac{c}{2}(c^2t^2-x^2)^{1/2}]}
\Theta(ct-x) .
\eeq
Linearizing the square root of the exponent to evaluate the short time 
Fourier transform of the saddle wave function (\ref{sr}),
one obtains 
\beq
S_s(t,\om;x)=\frac{2x^2}{\pi^2 T c^4 t^2 \theta}
\frac{\Os^2}{(\Os+\O0)^2}\frac{\sin^2[(\Om-\O0)T/2]}{(\Om-\Os)^2}\,\;
\;\;\;(t>T/2+x/c) .
\eeq
In this case the maximum of the envelop as a function of $t$
is given by 
\beq
t_{en}=\frac{\tau}{3^{1/2}}\left[1+3\left(\frac{\O0^2}{c^2/2}
\right)^2\right]^{1/2} .
\eeq
Because of the dispersion relation $\O0\le c^2/2$
in the evanescent case, it follows that 
$t_{en}$ is bounded by $\tau/3^{1/2}$ and $(4/3^{1/2}) \tau$.

\newpage

{\bf\large Figure Captions}
\vspace*{1cm}\\
\begin{enumerate}
\item \label{f1} 
Decimal logarithm of the amplified density,
$A=|\psi(x,t)|^2\exp(2\kappa_{0}x) $, versus
time for 
a signal emitted by a source located at $x=0$ with frequency
$\o0=0.5$, and observed at $x=7,14,21$
(solid, dotted, and dotted-dashed lines).
Noticeable for $x=7$ is the contribution of 
the saddle point solution $\psi_s$ (dashed line) in comparison 
with the exact solution $\psi$.
At $x=14,21$ the exact wave function $\psi$ and 
the saddle point solution $\psi_s$ are  
indistinguishable on the scale of the figure. 
\vspace*{.5cm}\\        

\item\label{f2}
Evolution of the average instantaneous frequency 
$\bar \om$, saddle point frequency $\os$, and signal frequency $\o0$
versus time (solid, dashed and 
dotted lines) for $x=7$ and $\o0=0.5$.
The circle marks the time $t_{tr}$ at which the pole contribution and saddle 
point contribution are equal in magnitude. 
\vspace*{.5cm}\\

\item\label{f3}
Decimal logarithm of the amplified density,
$A=|\psi(x,t)|^2\exp(2\kappa_{0}x)$, versus
time for $x=0.05$ and $x=0.45$ (solid and dashed thick lines respectively). 
The approximations provided by the saddle point solution $\psi_s$ are also 
shown (solid and dashed thin lines). $\o0=0.5$. The empty and filled squares 
mark the values of the traversal time $\tau$ for $x=0.05$ and $x=0.45$ 
respectively.
\vspace*{.5cm}\\

\item\label{f4}
Decimal logarithm of the amplified density,
$A=|\psi(x,t)|^2\exp(2\kappa_{0}x)$,
for the frequency band limited source and of the 
approximate solution $|D_+^0|^2$ versus
time (solid and dashed lines respectively), for $x=50$ (lower set), 
and $x=135$ (upper set).  
$\o0=0.5$, $\do=0.12$.   
\vspace*{.5cm}\\

\item\label{f5}
Decimal logarithm of the amplified density,
$A=|\psi(x,t)|^2\exp(2\kappa_{0}x)$,
of the 
approximations $|D_+|^2$ and $|D_+^0|^2$ versus
time (solid, dashed and dotted lines respectively), for $x=13.5$. 
The square marks the value of $\tau$.
$\o0=0.5$, $\do=0.12$ 
\vspace*{.5cm}\\

\item\label{f6}
Frequency evolution 
$\bar \om$, $\os$, and $\o0$ versus time (solid, dashed and 
dotted lines) of the frequency band limited source for $x=13.5$,
$\o0=0.5$, and $\do=0.12$
\vspace*{.5cm}\\

\item\label{f7}
Contour map of the spectrogram for $\o0=0.5$, $T=52.36$, $x=135$.
Notice the transition from the kinetic to the potential regime and 
the side wings around the main peak.   
\vspace*{.5cm}\\
 
\item\label{f8}
Spectrogram $S(t,\om=0,5)$ for  $\o0=0.5$, $T=52.36$, $x=135$.
The square marks the value of $\tau=95.4$. (This is the period that
corresponds to $\do=0.12$ in Figure D) $T_0=12.57$.    
\vspace*{.5cm}\\

\item\label{f9}
Spectrogram $S(t,\om=0,5)$ for $\o0=0.5$, $T=12.57$, $x=135$.
The square marks the value of $\tau=95.4$. $T_0=12.57$.
\end{enumerate}


\begin{references}
\bibitem{qm} The numerical value of $\om$ should not be overinterpreted. 
Only energy differences, and the differences between the corresponding
frequencies are  
invariant with respect to shifts of the origin and physically significant.
\bibitem{avw} Writting $\psi=|\psi|e^{i\phi}$ the local average
instantaneous frequency is   
defined as \cite{Cohenb}
%
$$
\bar{\om}(x)\equiv -\frac{\partial \phi(x,t)}{\partial t}\,.   
$$
%
Note that in this work the convention of signs for Fourier transforms 
and for $\bar{\om}(x)$ is 
different from the one in \cite{Cohenb}.
\end{references}
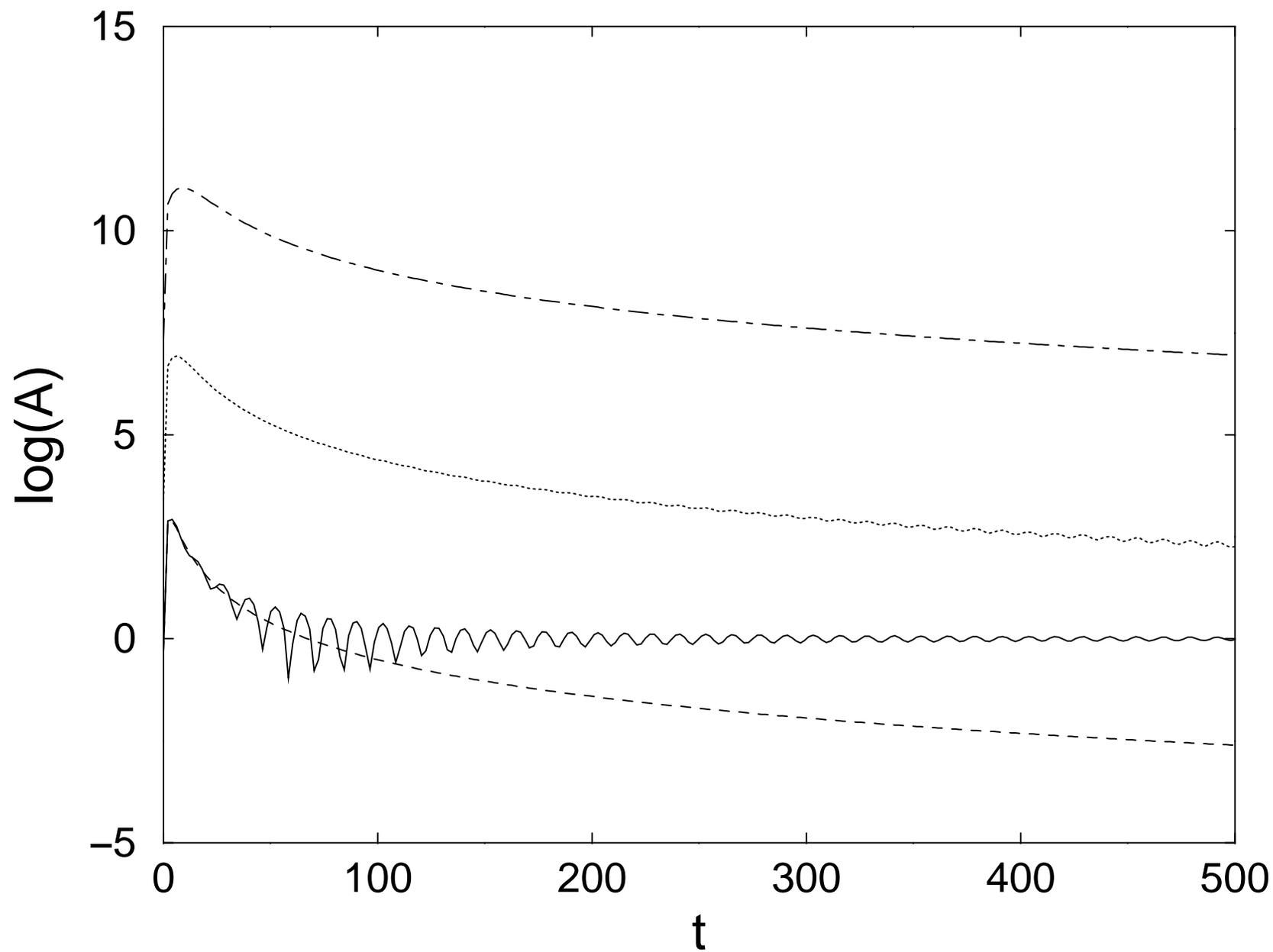
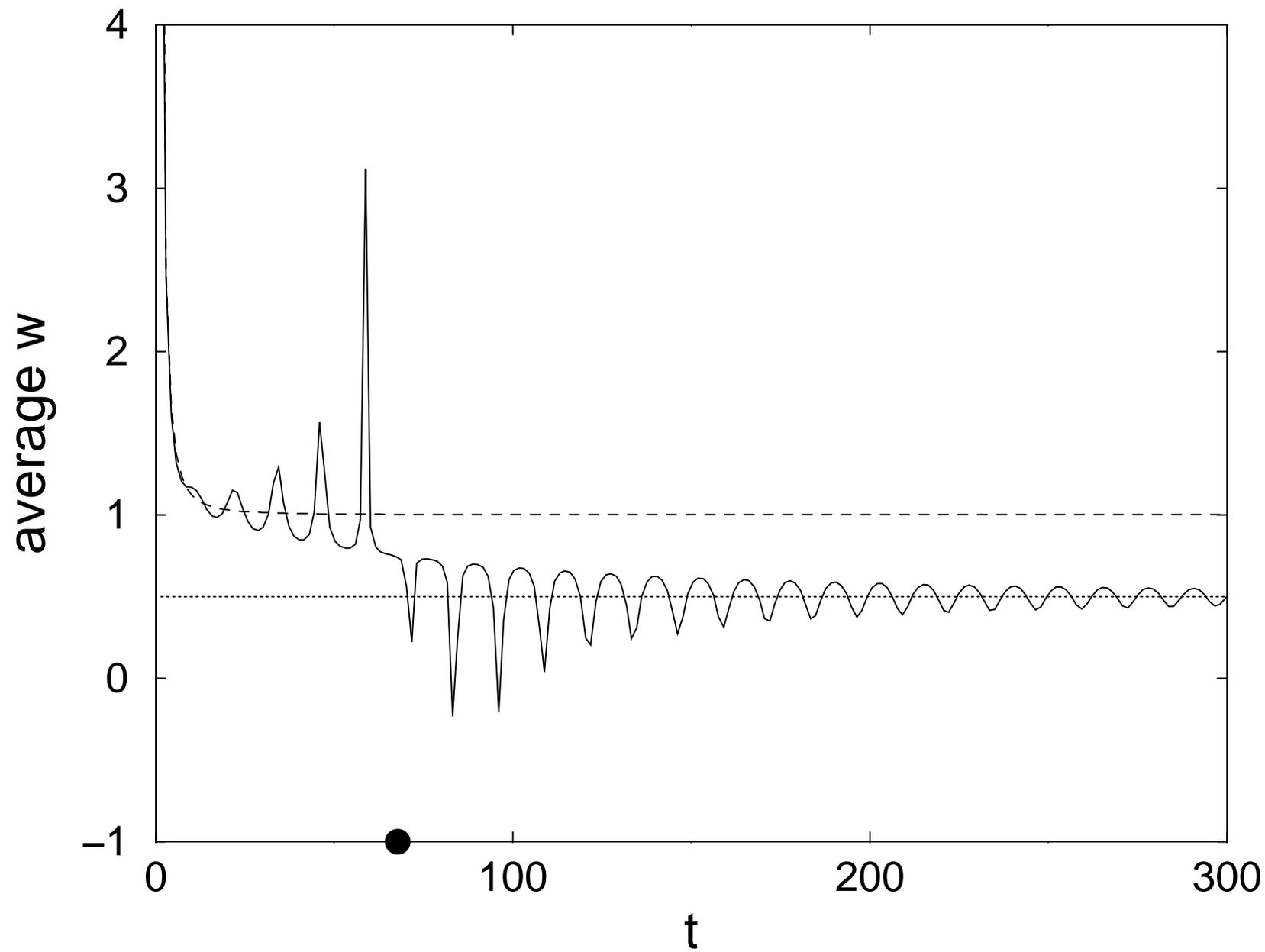
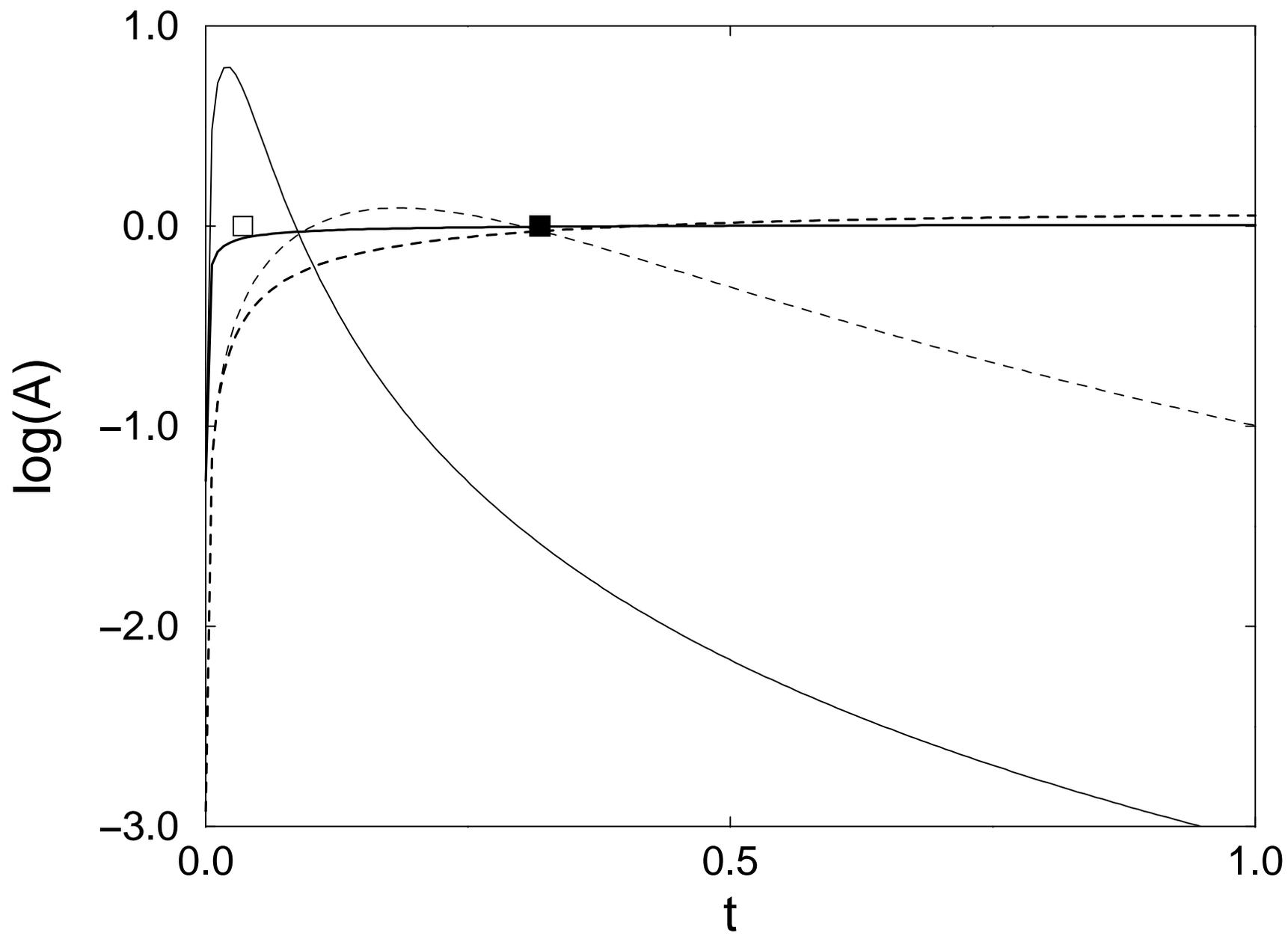
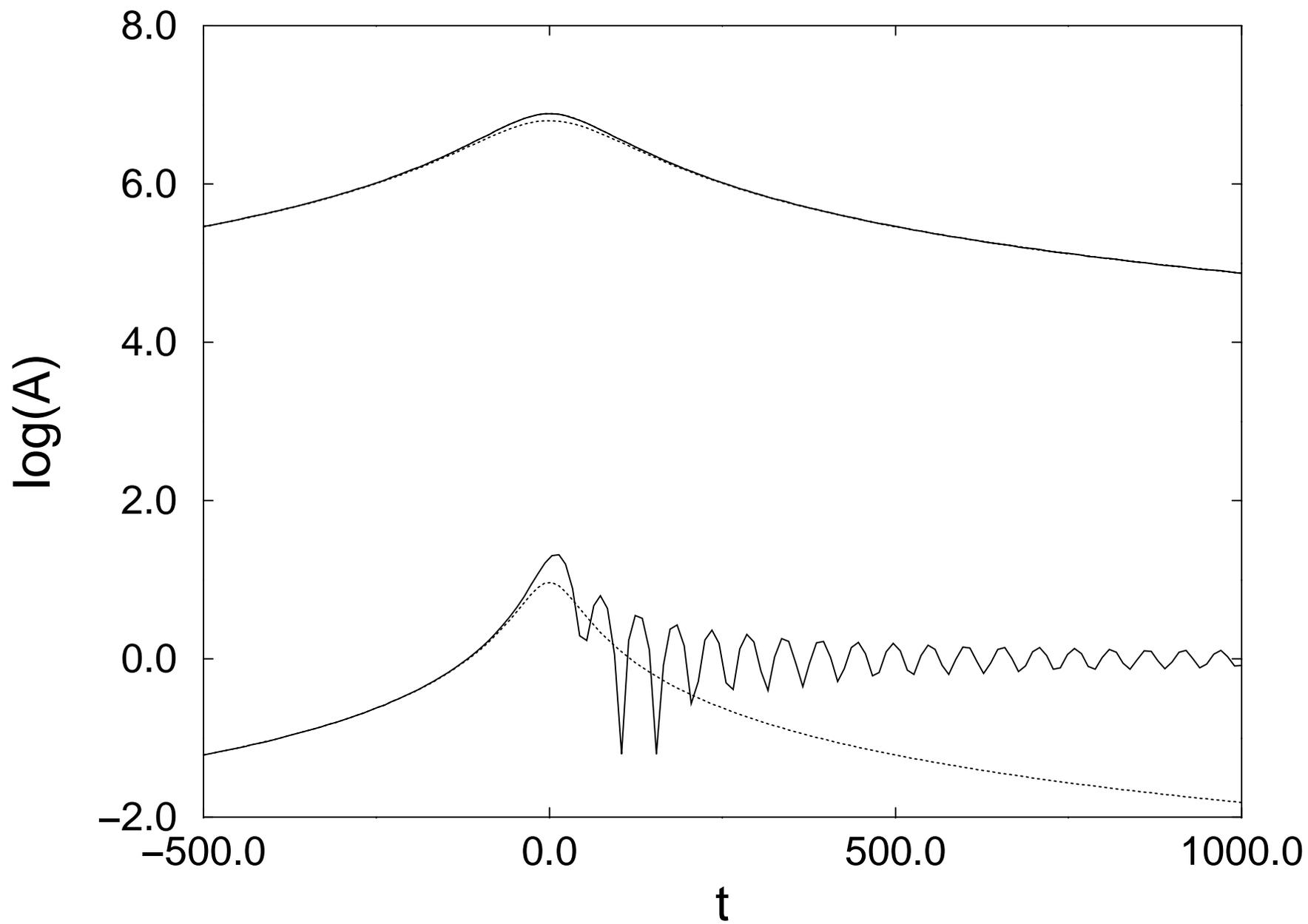
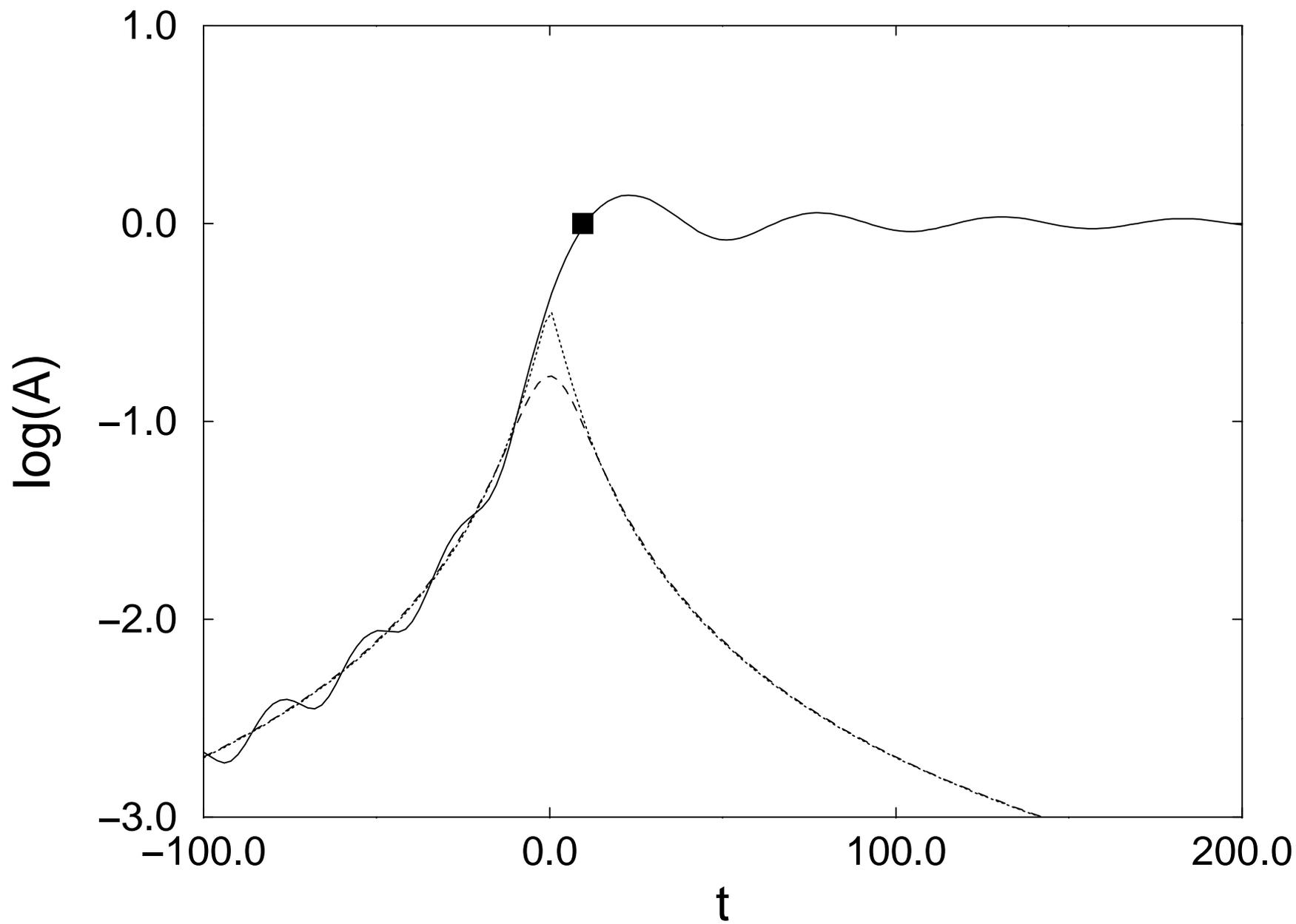
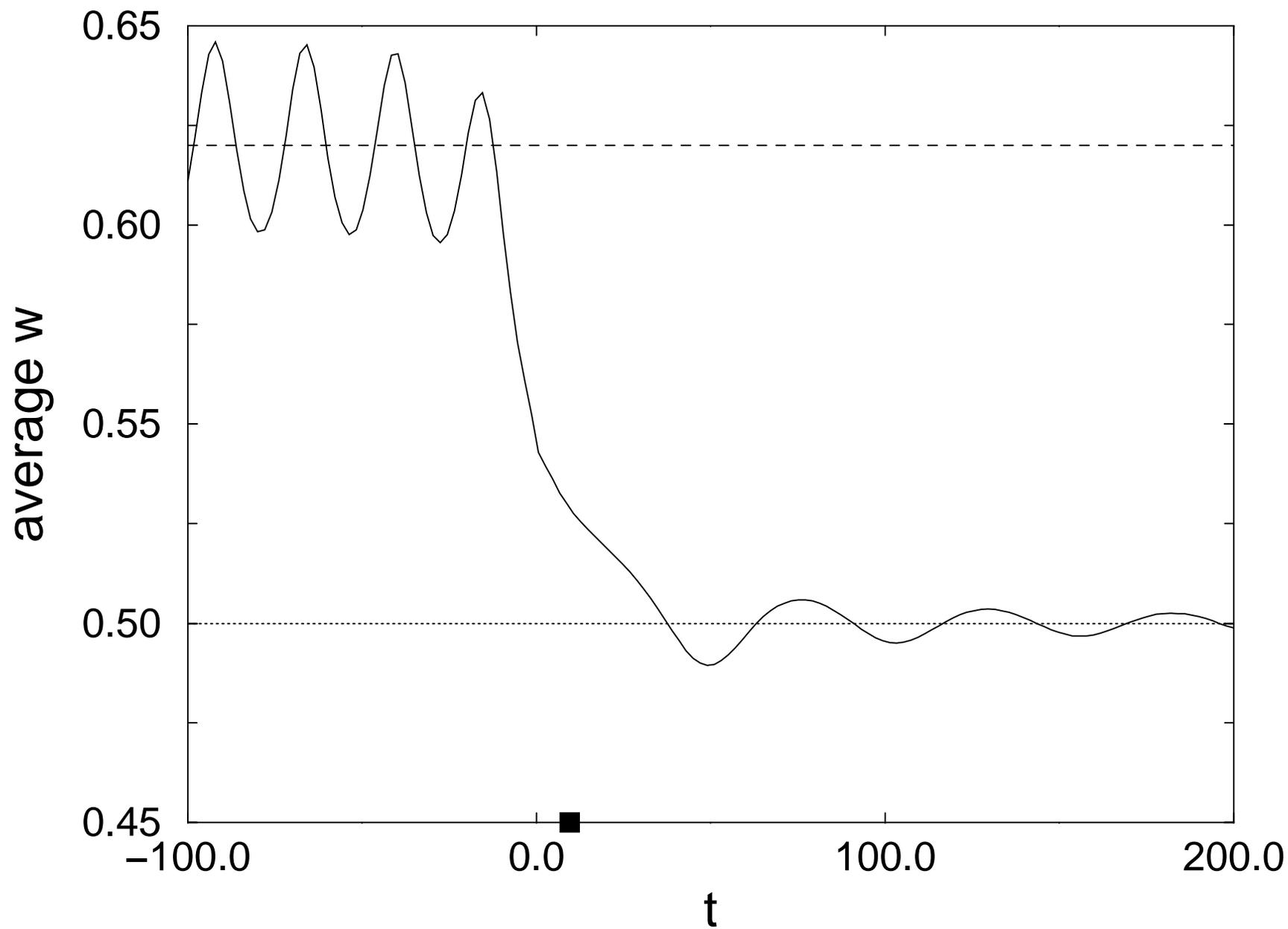
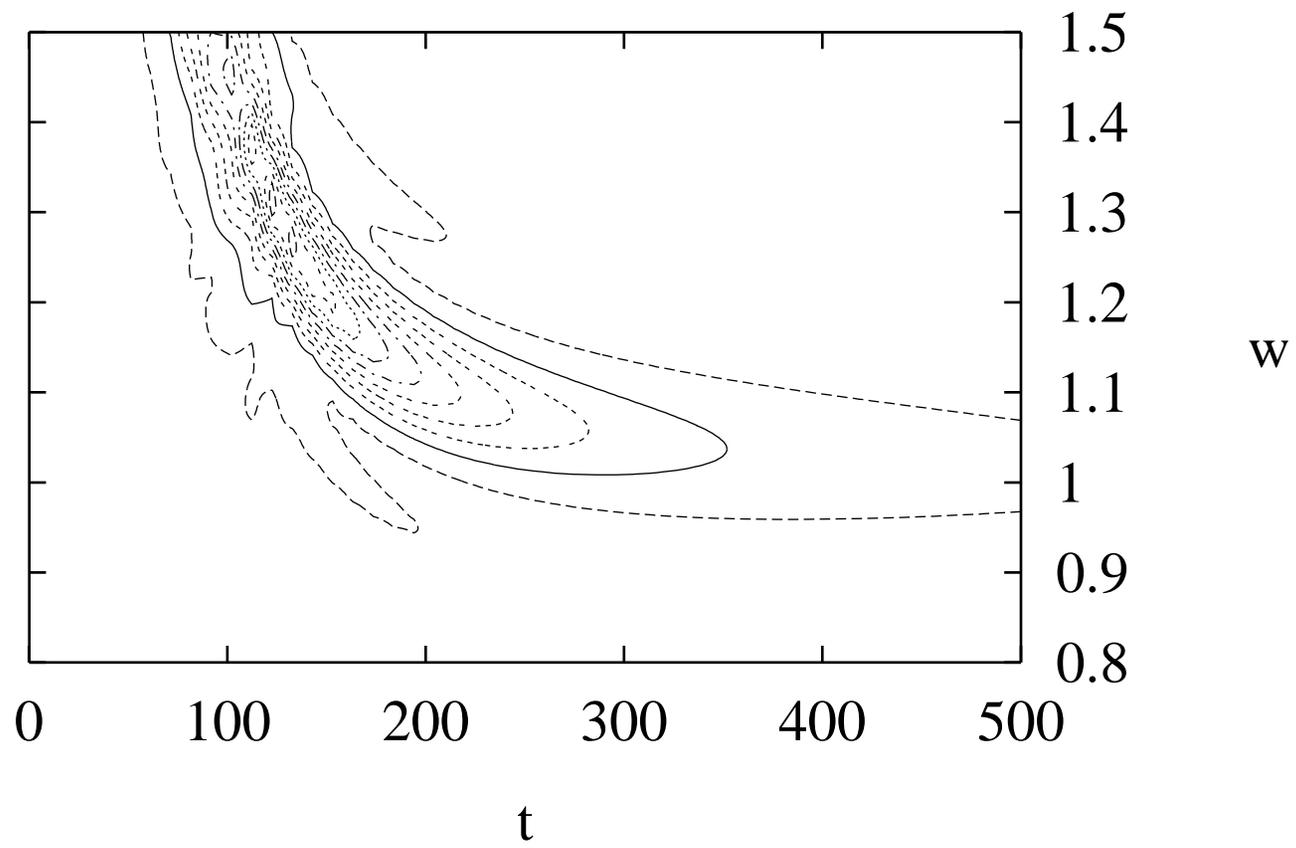
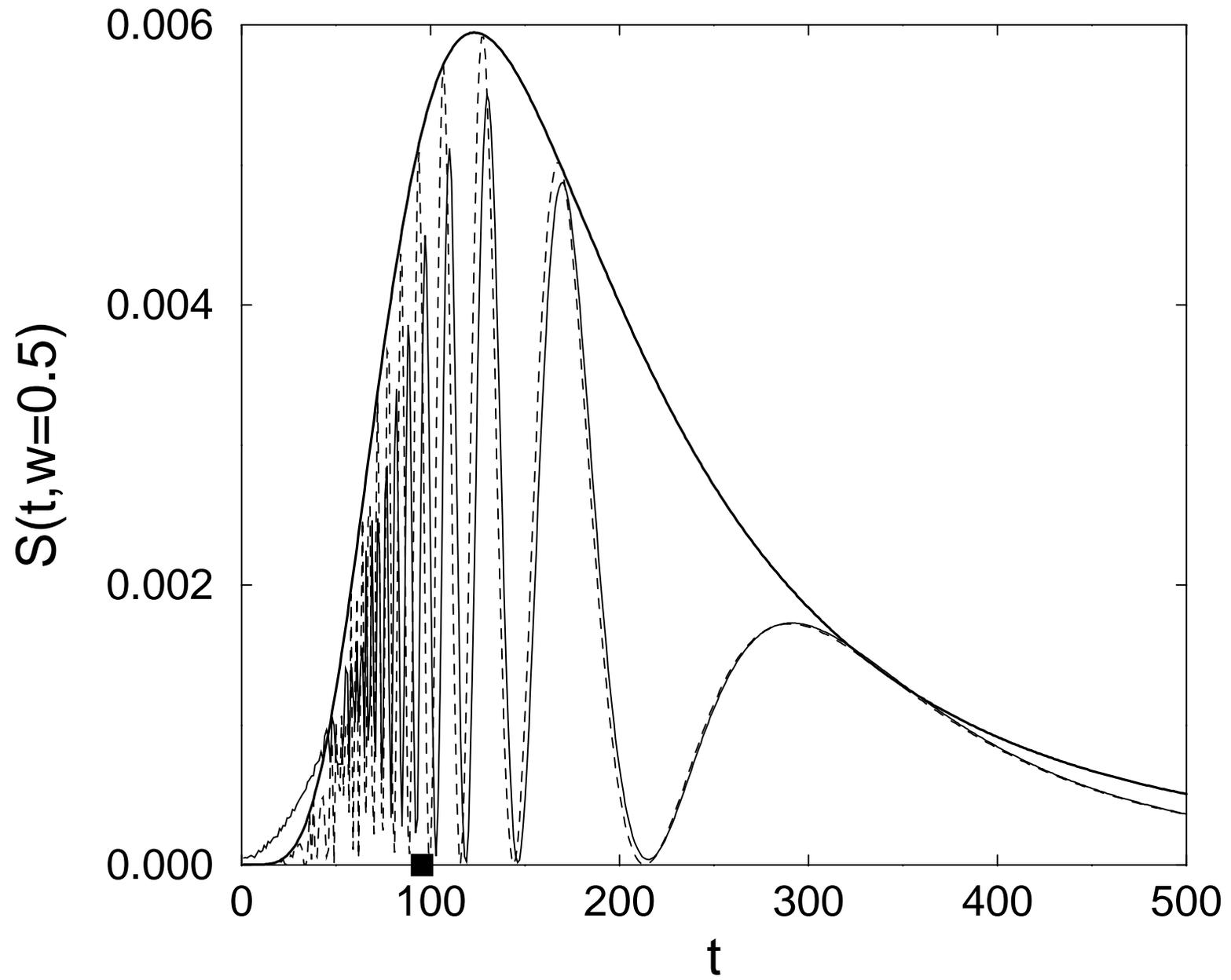
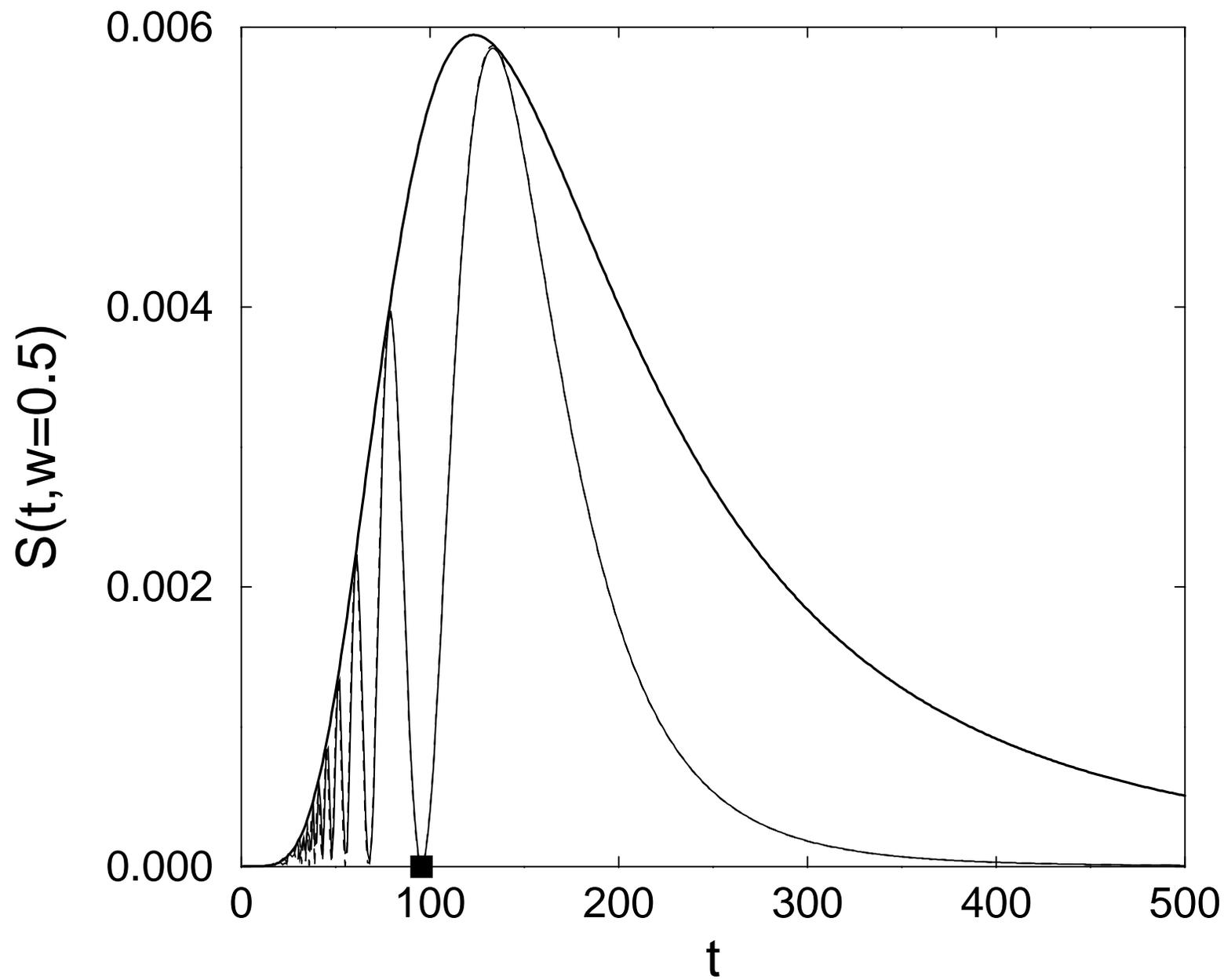
\end{document}